# Smart real-time system for heavy element research within new FLNR(JINR) Super Heavy Element Factory project: new approaches, algorithms, implementation.

**Yu.S.Tsyganov**[1], **A.N.Polyakov**[1], **V.A.Voinov**[1], **L.Schlattauer**[1,2]

*[1] Flerov Laboratory of Nuclear Reactions, Joint Institute for Nuclear Research, 6 Joliot-Curie, Dubna, Moscow region, 141980, Russia*

*[2] Department of Experimental Physics, Faculty of Science, Palacký University, 17. Listopadu 1192/12, 771 46 Olomouc, Czech Republic*

tyra@jinr.ru

***Abstract***

*The **D**ubna **G**as-**F**illed **R**ecoil **S**eparator is the most advanced facility currently in use in the field of research of Superheavy Nuclei (SHN)**.** During last year's, **IUPAC** established the priority of the **DGFRS** experiments in the discovery of new **Z=114-118** elements. Definitely, the **DGFRS** detection system and method of "active correlations" have played a significant role in these discoveries**.** Author defines abstract mathematical objects, like correlation graph and incoming event matrixes of a different nature in order to construct a simple procedure for detecting rare events, yet more exhaustive compared to the present one, using real-time detection mode. In this case one can use any of **n·(n-1)/2** correlation graphs to "trigger" cyclotron beam stopper to provide "background free" conditions to search for ultra-rare alpha decays. Here **n** is the number of correlation graph nodes. Schematics of these algorithms are presented.*

## 1. Introduction

During the past two decades the experimental attempts to reach the island of stability, i.e., nuclei at around the predicted Z=114 and N=184, have had great success resulting in the discoveries of elements with atomic numbers up to Z=118[1]. These nuclei were synthesized in complete fusion reactions of the double magic $^{48}$Ca (Z=20) used as a projectile and radioactive actinide targets from uranium to californium. Note, that all new elements Z=114-118 were synthesized firstly at the Dubna Gas-Filled Recoil Separator (DGFRS) [2]. As a result it become possible to produce SHN closer to the island of stability in $^{48}$Ca induced complete fusion reactions. The successes of the $^{48}$Ca-induced reactions were reasonable high evaporation residue cross sections observed for almost all cases.

Significant role in that successes have played “active correlation” (AC) method [3-6]. Namely with this technique it became possible to provide a deep suppression of background signals coming with the cyclotron beam. There is no negative influence to the quality of the experimental data if this method is used. That is why the AC method will be used in the framework of new FLNR DC-280 project, although with some modifications.

## 2. AC method – standard scenario.

The idea of the standard algorithm (method) is aimed at searching in real-time mode of time-energy-position recoil-alpha links, using the discrete representation of the sensitive layer of the position sensitive PIPS detector separately for events like “recoils” and “alpha-particle”[3]. So, the real PIPS detector is represented in the PC’s RAM in the form of two matrixes, one for the recoils and one for alpha-particles. Those elements are filled by values of elapsed times of the given elements. For the “alpha” elements the events have a composite nature. That is, the total energy of two components, one measured by the focal plane detector and one measured by any detector, surrounding the focal plane detector. In each case of “alpha-particle” detection, a comparison with “recoil-matrix” is made, involving neighboring elements (+/- 3 elements). If the minimum time is less or equal to the setting (fixed) time, the system turns the beam chopper which deflects the heavy ion beam in the injection line of the cyclotron for a short time (~1-2 min). The next step of the computer code ignores the vertical position of the forthcoming alpha-particle during the beam-off interval. If such decay takes place in the same strip that generated the pause, the duration of the beam-off interval is prolonged up to 10-30 min. The simplified block-diagram of the process is shown in Fig.1.

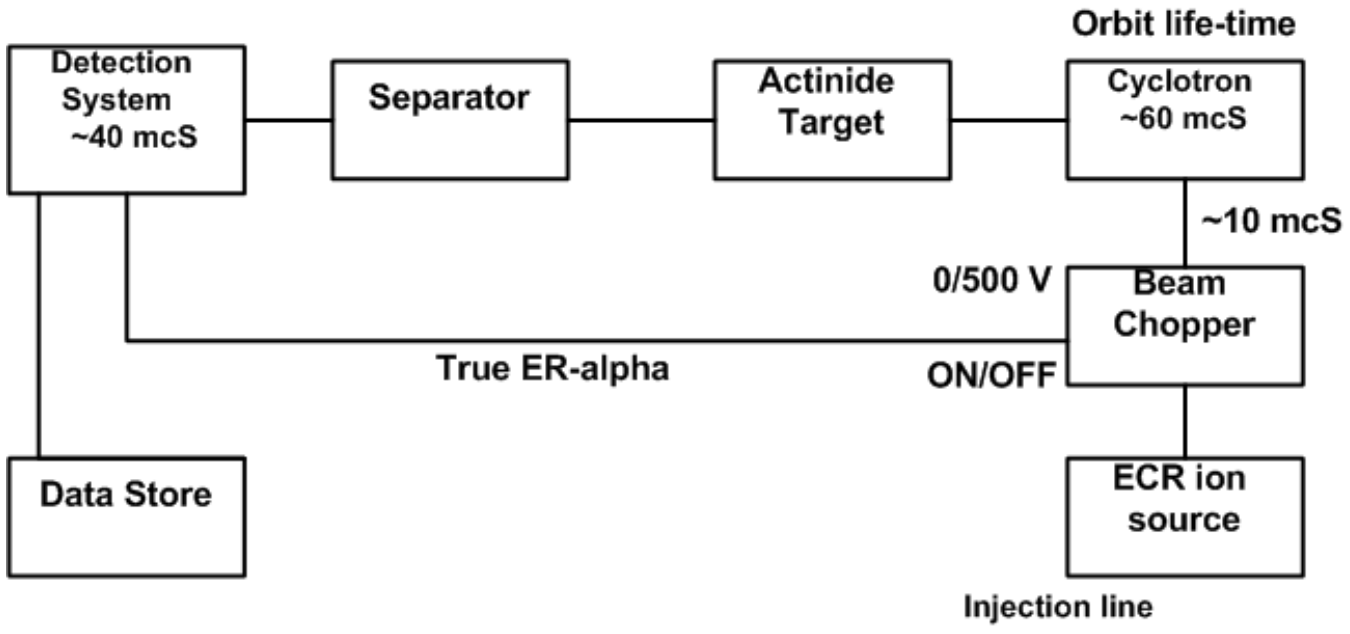

**Fig.1 The simplified block-diagram of the process**

### 3. A "flexible" scenario for a new real-time algorithm

The equation system for this approach is:

$$\eta(t) \le \varepsilon, \quad P_{dec}(\tau_{PS}) \ge 1-\xi, \quad \text{where } \varepsilon,\xi \ll 1 \text{ pre settled parameters.}$$

Here, $P_{dec}(t) = 1 - e^{-\lambda \cdot \tau_{PS}}$, where λ is a decay constant, $P_{dec}$-decay probability and $\eta$ – value of relative losses in the whole irradiation time value, $N$- effective pixel numbers of DSSSD detector (128x48), t- correlated time value, $\tau_{ps}$- pause time, $\nu_{ER,\alpha}$ – mean values of recoils or alpha event rate per element (pixel).

Therefore : $\eta = \frac{N}{2} \cdot P_1^{ER} \cdot P_1^{\alpha} \cdot \frac{\tau_{PS}}{t}$. Here: $P_1^{ER} = 1 - e^{-\nu_{ER} \cdot t}$ and $P_1^{\alpha}(t) = 1 - e^{-\nu_{\alpha} \cdot t}$ .

$P^{ER}$ *u* $P^{\alpha}$ – are random probabilities for recoils and alpha signals, respectively to be detected in the (0,t) time interval.

After transformation, taking into account rate number values for all DSSSD detector under condition of uniformity, one can rewrite: $F_{ER} = N \cdot \nu_{ER}$ и $F_{\alpha} = N \cdot \nu_{\alpha}$ and respectively: $P_1^{ER}(t) = 1 - e^{-\frac{F_{ER} \cdot t}{N}}$, $P_1^{\alpha}(t) = 1 - e^{-\frac{F_{\alpha} \cdot t}{N}}$ и $\eta = \frac{\tau_{PS}}{2t} \cdot (1 - e^{-\frac{F_{ER} \cdot t}{N}}) \cdot (1 - e^{-\frac{F_{\alpha} \cdot t}{N}})$.

In the Fig.2 the flowchart of the ResdStorm#2 C++ code is shown. The subroutines which correspond to the "flexibility" of the scenario are marked in grey.

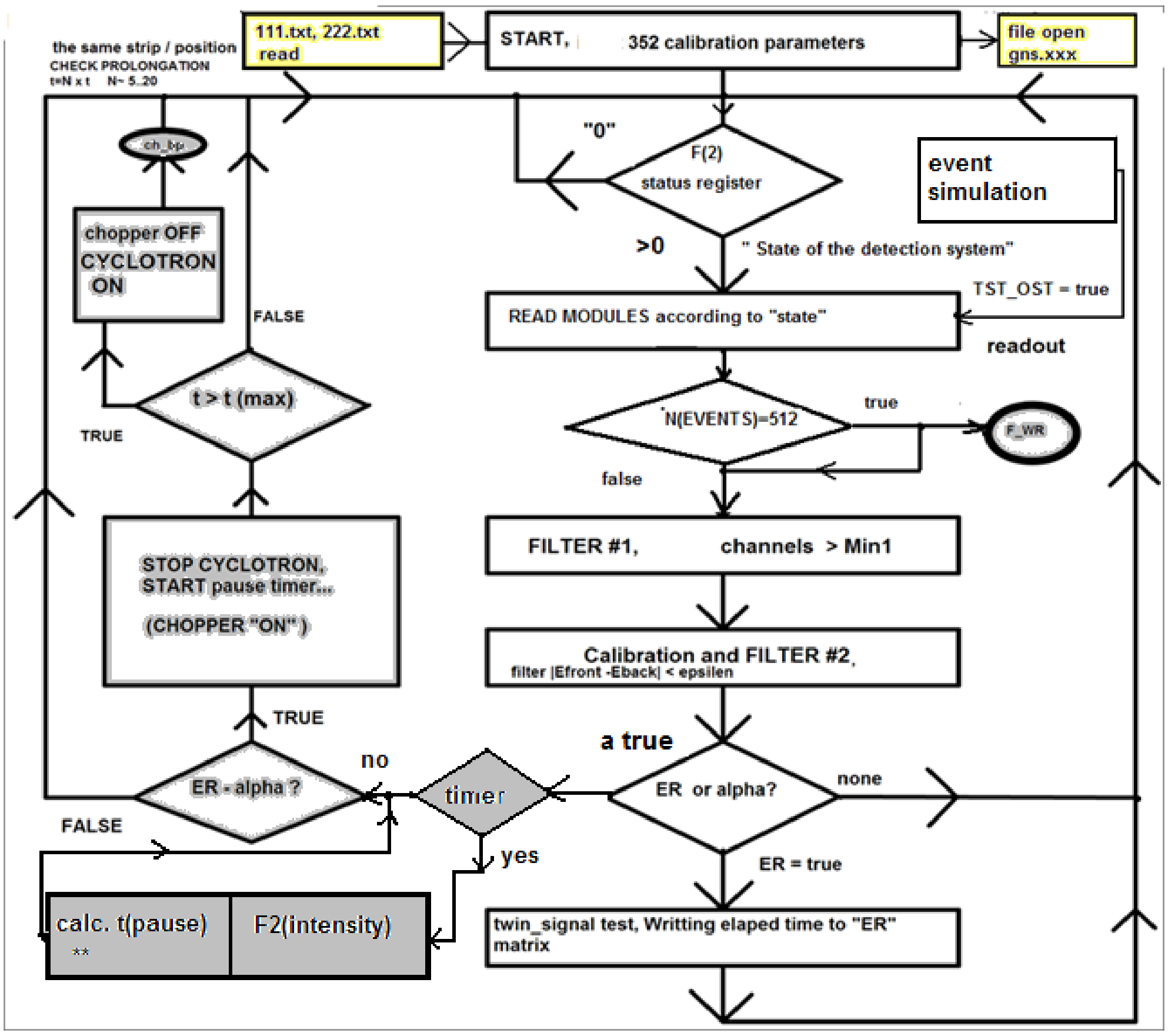

**Fig.2 Flowchart of C++ RedStorm#2 code**

Another way to apply "flexible" scenario is to use directly elapsed beam stops number value as a main experimental parameter to change correlation time interval or beam pause time interval ("natural", trivial scenario). A drawback of this approach is in very low statistics of the beam stops number value for a time of minutes- tens of minutes. It means that very long time is required to make next iteration to change parameters. From the other hands, evident advantage of this algorithm is in its simplicity. If we use the mentioned algorithm, each *n+1* iteration equation can be written as: $\tau_{PS}^{n+1} \approx \tau_{PS}^{n} \cdot \frac{\eta_{fix}}{\eta_{meas}^{n}} \cdot \left(\frac{< J^{fix} >}{J_n}\right)^2$ **[7,8].**

In this formula: $\tau_{PS}$ – beam pause time interval, $\eta$- relative value of irradiation time value (*fixed and measured, respectively*), *J*- beam intensity measured with Faraday cup, *n*-iteration number value.

## 4. Summary

During the past two decades the “active correlations” method was successfully used to provide radical background suppression in heavy-ion induced complete fusion reactions. This method is using several pre-settled parameters for a calculation. We created a new version of real-time algorithm for the case of using more “flexible” scenario. Developed code called C++ Redstorm#2 PC autonomously decides how long the correlation will last and determine the pause time intervals after each beam stop event.